\def\ref{\par\noindent\hang}

\def\spose#1{\hbox to 0pt{#1\hss}}
\def\approxlt{\mathrel{\spose{\lower 3pt\hbox{$\sim$}}
	\raise 2.0pt\hbox{$<$}}}
\def\approxgt{\mathrel{\spose{\lower 3pt\hbox{$\sim$}}
	\raise 2.0pt\hbox{$>$}}}

\def\multleft#1{\hbox to size{\vbox {\halign {\lft{##}\cr #1}}\hfill}\par}
\def\multright#1{\hbox to size{\vbox {\halign {\rt{##}\cr #1}}\hfill}\par}

\def\today{\ifcase\month\or January\or February\or March\or April\or May\or
      June\or July\or August\or September\or October\or November\or December\fi
      \space\number\day, \number\year}
\def\$<${\thinspace}
\def\s{\hbox{\phantom{5}}}	

\def\boxit#1{\vbox{\hrule\hbox{\vrule\kern3pt\vbox{\kern3pt
          #1 \kern3pt}\kern3pt\vrule}\hrule}}

\def\cm{{\rm\thinspace cm}}

\def\erg{{\rm\thinspace erg}}

\def\keV{{\rm\thinspace keV}}
\def\km{{\rm\thinspace km}}

\def\Mpc{{\rm\thinspace Mpc}}

\def\s{{\rm\thinspace s}}

\def\Hz{{\rm\thinspace Hz}}

\def\ergpcmsqps{\hbox{$\erg\cm^{-2}\s^{-1}\,$}}

\def\ergps{\hbox{$\erg\s^{-1}\,$}}

\def\kmps{\hbox{$\km\s^{-1}\,$}}

\def\psqcm{\hbox{$\cm^{-2}\,$}}

\def\kmpspMpc{\hbox{$\kmps\Mpc^{-1}$}}

\documentstyle[psfig]{mn}
\begin{document}
\hsize=6truein

\title{The ASCA spectrum of the z=4.72 blazar, GB
1428+4217}

\author[]
{\parbox[]{6.in} {A.C.~Fabian$^1$, K. Iwasawa$^1$, A. Celotti$^{1,2}$, 
W.N.~Brandt$^3$, R.G. McMahon$^1$ and  I.M. Hook$^4$ \\
\footnotesize
1. Institute of Astronomy, Madingley Road, Cambridge CB3 0HA \\
2. S.I.S.S.A., via Beirut 2-4, 34014 Trieste, Italy\\ 
3. Harvard-Smithsonian Center for Astrophysics, 60 Garden Street,
Cambridge, MA 02138 USA\\
4. European Southern Observatory, D-85748 Garching, Germany\\}}
\maketitle

\begin{abstract}
The X-ray luminous quasar GB 1428+4217 at redshift 4.72 has been observed
with ASCA. The observed 0.5--10~keV flux is $3.2\times 10^{-12}\ergpcmsqps$.
We report here on the intrinsic $4-57\keV$ X-ray spectrum, which is very
flat (photon index of 1.29).  We find no evidence for flux variability
within the ASCA dataset or between it and ROSAT data. We show that the
overall spectral energy distribution of GB 1428+4217 is similar to that of
lower redshift MeV blazars and present models which fit the available data.
The Doppler beaming factor is likely to be at least 8. We speculate on the
number density of such high redshift blazars, which must contain
rapidly-formed massive black holes.

\end{abstract}


\section{Introduction}

The distant radio-loud quasar GB~1428+4217 at redshift $z=4.72$ (Hook \&
McMahon 1997) is so X-ray bright that its X-ray emission dominates the
observed spectral energy distribution (Fabian et al 1997). This property,
together with its flat-spectrum, compact, radio appearance (Patnaik et al
1992) suggest that it is a blazar which is beamed toward us.

The ROSAT flux from GB~1428+4217, after a small correction for Galactic
absorption, corresponds to $\sim 10^{-12}\ergpcmsqps$ in the observed
0.1--2.4~keV band, which is bright enough for a good spectrum to be
obtained with ASCA. We report here on that spectrum, which is flat and
indicates that the overall spectrum peaks in the hard X-ray or gamma-ray
bands. The flux of GB~1428+4217 in the 0.5--10~keV ASCA band is $\sim
3.2\times 10^{-12}\ergpcmsqps$, more than three times brighter than the
next brightest quasar at $z>4$, GB~1508+5714 (Hook et al 1994; Moran \&
Helfand 1997) which has similar properties to GB~1428+4217.

\section{The ASCA spectrum}

GB~1428+4217 was observed with ASCA (Tanaka, Inoue \& Holt 1994) on 1997
January 17. The Solid state Imaging Spectrometers (SIS; S0 and S1) were
operated in Faint mode through the observation using the single standard
CCD chip for each detector. The lower level discriminator was set at 0.47
keV for the SIS. The spectral resolution of the SIS at the time of the
observation had FWHM$\sim$ 300 eV at 6 keV due to degradation in the
performance of the CCDs (Dotani et al. 1997). Nominal PH mode was used
for the Gas Imaging Spectrometers (GIS; G2 and G3).

Standard calibration and data reduction methods were employed using
FTOOLS provided by the ASCA Guest Observer Facility at Goddard Space
Flight Center. The net exposure time was 40.2~ks for each detector.
During the one day observation, no significant flux variation (greater
than 30 per cent) was observed. Background data were taken from a blank
part of the field in the same detector. Spectral fits were performed
jointly for the background-subtracted data from all detectors using
XSPEC. The normalization of the spectra for the SIS and GIS detectors was
allowed to be slightly (12 per cent) different (see Gendreau \& Yaqoob 1997
for a discussion of ASCA calibration issues).

The data are well-fit by a hard power-law spectrum (Fig.~1). Assuming
only the Galactic column density of $1.4\times 10^{20}\psqcm$ (Dickey \&
Lockman 1990), the photon index $\Gamma=1.29\pm0.05$ ($\chi^2/{\rm
dof}=383/380$). A 90 per cent confidence upper limit to the column
density is $5.3\times 10^{20}\psqcm$ in the observed frame or $4.5\times
10^{22}\psqcm$ at $z=4.72$. The limit on a joint fit of $\Gamma$ and
intrinsic absorption $\Delta N_{\rm H}$ is shown in Fig.~2. There is no
obvious break seen in the spectrum (see Fig.~1); a broken power-law fit
to the data constrains the allowable difference in spectral index to be 
$\pm0.1$ at 3~keV, increasing to $\pm0.2$ at 1.7~keV and greater than
$\pm0.4$ below 1.2~keV. 

We note that the SIS data alone do suggest the presence of some excess
absorption, corresponding to $\Delta N_{\rm H}=1.3\pm0.6\times
10^{23}\psqcm$ at $z=4.72$. The photon index is then $\Gamma=1.43\pm0.12$.

\begin{figure}
\centerline{\psfig{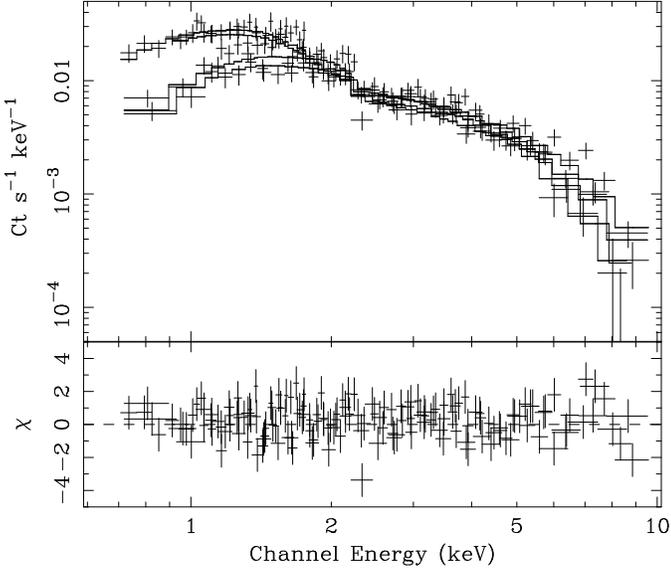}}
\caption{ASCA spectrum for GB~1428+4217. The SIS and GIS spectra are the upper
and lower ones below 2~keV, respectively.}
\end{figure}

\begin{figure}
\centerline{\psfig{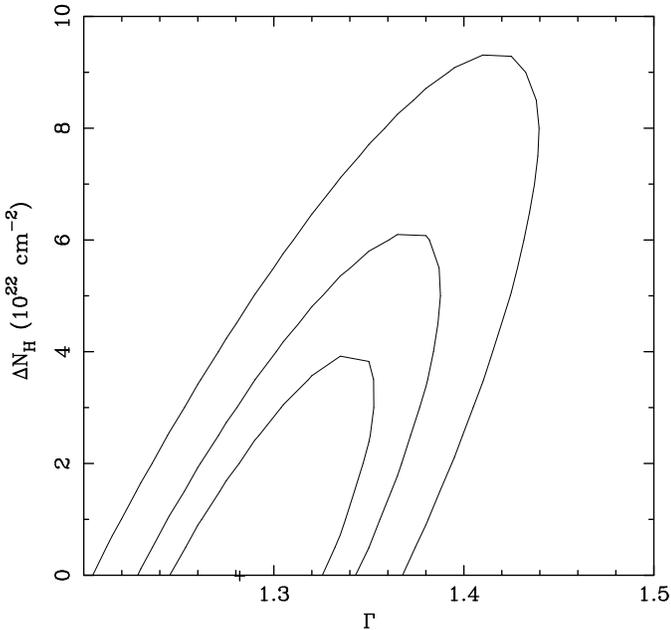}}
\caption{Excess absorption column density $\Delta N_{\rm H}$ (at $z=4.72$)
plotted against photon index for the joint SIS and GIS fits. The contours
correspond to confidence levels of 68, 90 and 99 per cent. Note that cosmic
abundances are assumed, which may not be relevant for such a young object.}
\end{figure}

No obvious emission or absorption features due to iron are evident in the
spectrum. The 90 per cent confidence limit on the equivalent width of a
narrow emission line from either cold (6.4~keV) or ionized (6.7~keV) iron is
about 17~eV or just under 100~eV in the quasar frame. This last limit rises
to about 170~eV if the line width has a dispersion of 0.5~keV, instead of
0.1~keV. Some marginal structure in the SIS spectrum is observed around
1~keV in the observed frame ($\sim 5.9\keV$ in the rest frame of the quasar).

The flux of GB~1428+4217 in our 2--10~keV band (after correction for
Galactic absorption and using the best-fitting model) is $2.5\times
10^{-12}\ergpcmsqps$ and in the 0.1--2.4~keV ROSAT band is $1.29\times
10^{-12}\ergpcmsqps$. These translate to a rest-frame, 2--10~keV luminosity
of $1.4\times 10^{47}\ergps$ (assuming $H_0=50\kmpspMpc$ and $q_0=0.5$),
rising to $6.4\times 10^{47}\ergps$ in the rest-frame 4-57~keV band which we
observed. The ASCA flux in the 0.1--2.4~keV ROSAT band is consistent, within
the uncertainties, with that detected during the pointed ROSAT observations.
The flux reported from the ROSAT All-Sky Survey (see Table 2 of
unidentified sources in Brinkmann et al 1997) is marginally weaker.

\section{An MeV Blazar?}

Here we consider the ASCA results in the wider context of the broad--band
energy distribution and examine possible interpretations of the
properties of this source. 

As discussed by Fabian et al. (1997), the flat X--ray and radio
spectra, the remarkable spectral energy distribution (SED) peaking
in the X-ray band, and the radio polarization of GB~1428+4217 all
suggest that it is a blazar candidate, the observed emission of which is
therefore dominated by relativistic beaming.  Although, due to the
sparse data coverage, this cannot be established uniquely, we consider
it to be the most likely possibility and therefore interpret here the
properties of GB~1428+4217 within this scenario.

In particular the flat radio spectrum indicates that the source is a
potentially strong $\gamma$--ray emitter. Blazar SEDs are characterized
by two broad components, peaking in the IR--UV and $\gamma$--ray band,
most likely produced as synchrotron and inverse Compton emission,
respectively. It has also been recognized, through a study of the
spectral indices of the $\gamma$--ray loud sample (Comastri et al. 1997),
that the position and the relative intensities of these two peaks are
correlated (see also Ghisellini et al. 1997, in prep).

The X--ray spectrum of GB~1428+4217 is among the flattest measured in
$\gamma$--ray loud blazars, suggesting that the source emission peaks
in the inverse Compton component at $\sim$ MeV energies. If one
proceeds by considering the similarities with lower redshift sources,
the seed photons for the Comptonization are most likely provided, 
at least in radio--loud quasars for which there is strong evidence of
an intense quasi--thermal radiation field, by photons produced
externally to the relativistically moving plasma. Indeed the
extremely flat IR-optical spectrum (see Fig.~3) could represent the
seed photon field.

The (non--simultaneous) broad band distribution (in $\nu L_{\nu}$) is shown
in Fig.~3, where, together with the radio (see Fabian et al. 1997),
IR--optical (Hook \& McMahon 1997; Hoenig et al in preparation), ROSAT and
ASCA spectra, IRAS and CGRO/EGRET upper limits are also plotted. In
particular GB~1428+4217 has not been detected by EGRET (Thompson et al
1995); the $\gamma$--ray limit reported in the figure corresponds to a
sensitivity level of $\sim 10^{-7}$ ph cm$^{-2}$ s$^{-1}$ above 100 MeV.

\begin{figure*}
\centerline{\psfig{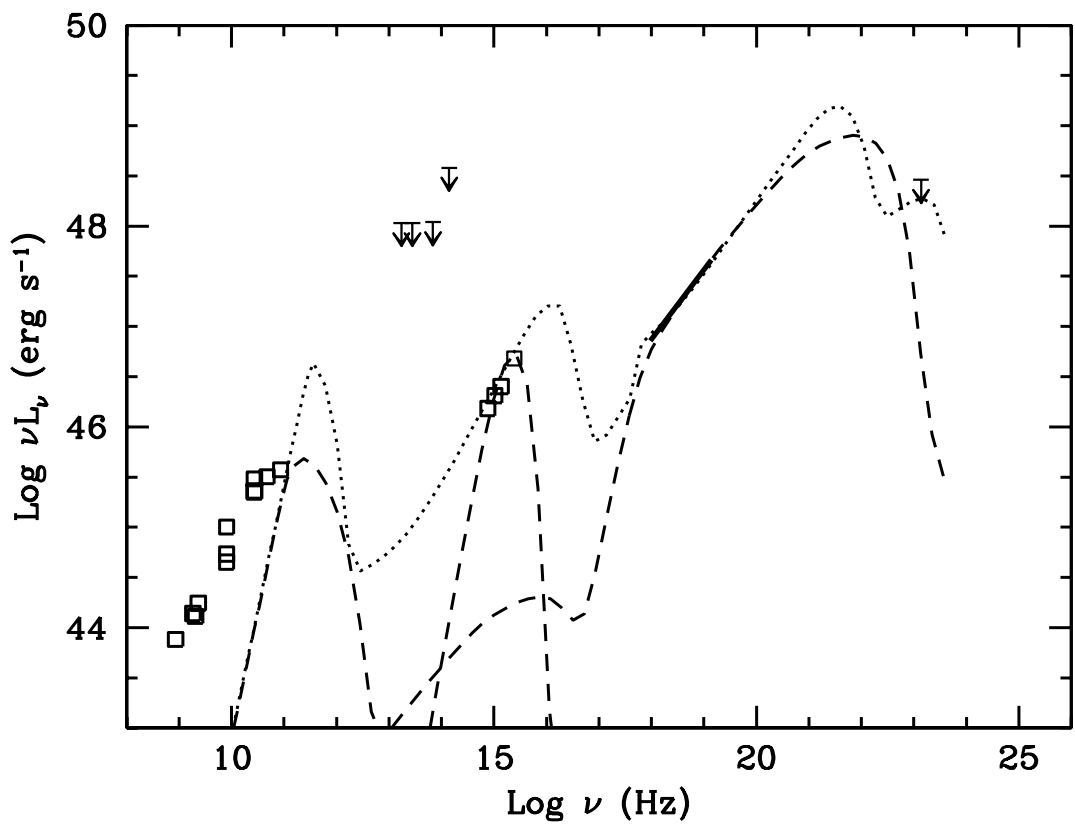}}
\caption{Broadband SED (rest frame in $\nu L_{\nu}$) with possible
blazar model spectra. The three peaks in the continua are due to: a)
$\sim 10^{11}\Hz$, self-absorbed synchrotron radiation; b) $\sim
10^{16}\Hz$, first order SSC radiation; and c) $\sim 10^{22}\Hz$,
Comptonization of the radiation field dominating in the optical/UV band.
This field is either external (EC, blackbody) or internal (SSC)
radiation, or some combination of them (SSC/EC). Shown here is an EC
model (dashed), with the external blackbody radiation field peaking in
the UV (it can of course peak at slightly higher frequencies), and an
SSC/EC model (dotted). The parameters of these and a pure SSC model are
given in Table 1. }
\end{figure*}

Despite the sparse coverage of the total SED of GB~1428+4217, we now
examine the predictions of plausible emission models.

\subsection{Emission models}

In the simplest hypothesis, the emission region is assumed to be a
homogeneous sphere, moving with bulk velocity $\beta c$ along a direction
close to the line of sight. The region, of dimension $R$, is filled with
a tangled magnetic field of intensity $B$. Here relativistic electrons
are continuously injected with a power-law spectrum extending in energy
to $\gamma_{\rm max} m_{\rm e} c^2$, at a rate per unit volume
corresponding to an injection compactness parameter $\ell_{\rm inj}$.

The (stationary) equilibrium particle distribution is obtained by solving
the continuity equation, balancing the rates of particle injection,
radiative cooling and electron--positron pair production through
photon--photon collisions.  In particular the cooling is dominated by the
synchrotron and inverse Compton processes, where the latter allows for
contribution to the radiation field both by photons produced inside the
emitting region as well as any external photon field (described for
simplicity as blackbody radiation at optical--UV frequencies).  Further
details on the precise assumptions of the code can be found in the paper
by Ghisellini et al. (1997, in prep.; see also Ghisellini 1997).

A complete exploration of the possible parameter space of the model is
beyond the aim of this paper. In Fig.~3 we superpose on the SED of
GB~1428+4217 the predictions of this homogeneous model. In particular
the broad band energy distribution represented by the dashed line has
been computed by assuming that the external photon field mainly
contributes the seed photons scattered to high (X-- and $\gamma$--ray)
energies through the inverse Compton effect. This external field could
also be observable in the optical band, as shown in Fig.~3
(dashed blackbody component peaking at $\sim 10^{15}$ Hz).

Indeed, the extremely flat near IR--optical spectrum cannot be simply
interpreted as synchrotron emission, as is usually considered (e.g.
Ghisellini 1997). Here we examine an alternative possibility, namely that
the near IR--optical radiation is dominated by SSC emission. The low
equivalent width of the UV emission lines (Hook \& McMahon 1997) supports
the presence of the boosted UV continuum then predicted. The
X-ray/$\gamma$-ray component can now be dominated by the Comptonization
of either this same SSC radiation or an external field. The SED predicted
under the last hypothesis (i.e. a hybrid SSC/EC model) is shown by the
dotted line in Fig.~3; a satisfactory model can also be found for the
pure SSC case. The parameters for all these models are reported in
Table~1.

\begin{table}
\caption{The input parameters for the EC, SSC/EC and SSC models:
(1) Model; (2), (3) compactnesses in injected particles and external
radiation field, respectively; (4) maximum energy of the injected
particles; (5) magnetic field intensity (G); (6) relativistic
Doppler factor.}
\begin{tabular}{@{}lcccccc}
\hline Model  &$\ell_{\rm inj}$ &$\ell_{\rm ext}$
&$\gamma_{\rm max}$  &$B$   &$\delta$\\
(1) & (2) &(3) & (4) & (5)& (6) \\
\hline

EC  &0.1   &1     &400 &0.1 &12     \\
SSC/EC &0.3   &4.5e-3&250 &0.1 &22    \\
SSC  &0.5  &0.  &350  &0.03  &22  \\

\hline
\end{tabular}
\end{table}

If our interpretation of GB~1428+4217 is correct two main points have to
be stressed: i) the set of parameters adopted here are consistent with
those analogously derived for a sample of $\gamma$--ray loud blazars by
Ghisellini et al. (1997, in prep), no evidence has been found so far of
any peculiar properties of this $z=4.7$ source compared to the low
redshift counterparts; ii) the observed luminosity of the source is
expected to be completely dominated by high energy radiation and copious
$\gamma$--ray emission, at a level corresponding to an isotropic
luminosity of 10$^{49}$ erg s$^{-1}$.

We also considered the predictions of an inhomogeneous SSC model
(Ghisellini \& Maraschi 1989) which can effectively account for the
X--ray and radio emission (produced in the less compact region of the
jet).  However, as already pointed out, the near IR--optical
spectrum cannot be reproduced as synchrotron emission and has then to
be ascribed to a different spectral component.

\subsection{Cosmological and statistical consequences}

If GB~1428+4217 is an MeV blazar, independently of the details of
the emission model, it is possible to derive general constraints by
assuming a typical dimension for the emitting source. In fact, the
requirement of transparency to internal photon--photon absorption (to the
process of pair production) implies $\delta \approxgt 8$, for $R\sim
2\times 10^{16}$ cm.  This dimension is similar to those of lower
redshift objects, and corresponds to an observed variability timescale of
$t_{\rm var}\sim 6$ days ($\delta\propto t_{\rm var}^{0.22}$).

Even assuming that the radiation observed from GB~1428+4217 is
amplified by relativistic beaming with, say, a bulk Doppler factor of
$\delta\sim 10
\delta_{1}$, the intrinsic source luminosity would amount to about
$10^{45}
\delta_{1}^{-4}$ erg s$^{-1}$, requiring a $\sim 10^7 M_{\sun}$ black
hole radiating at the Eddington rate being formed in less than a
billion years.

$\gamma$--ray emission from high redshift sources can be a very
powerful tool to study the background radiation fields. However, if this
source is indeed an ``MeV blazar'', despite the intense emission and great 
distance,  we do not expect to detect signs of
absorption due to the optical--UV stellar background radiation, as this 
mostly affects photons at energies $\approxgt $ GeV (Salamon \&
Stecker 1997).

The number density of sources at $z\approxgt 4$ is clearly critical to
the determination of the evolution of active nuclei (and the radio
phenomenon) and the inference of constraints on the formation or cycle of
activity of these systems. A minimum comoving number density has been
estimated from the three $z>4$ radio-loud quasars discovered by Hook \&
McMahon (1997). Alternatively, one can consider the number of sources
expected by statistical arguments based on the random orientation of the
beaming pattern in the Sky to obtain a maximum number. If objects are
classified as flat spectrum, radio-loud quasars for angles with the line
of sight smaller than $\approxlt 14^o$ (e.g.  Padovani \& Urry 1992),
then given at least one within a search area, about 6$\delta^2_{1}$ such
quasars are expected in total (with different luminosities). For the
assumed cosmological parameters, the resulting comoving number density of
very powerful flat spectrum, radio quasars is $1.2-4\times 10^{-10}$
Mpc$^{-3}$ at $4<z<5$.

Many more objects ($\sim 200\delta_1^2$) which are not beamed in our
direction at $4<z<5$ are predicted to be in the search area. They may
appear as powerful radio galaxies and steep spectrum radio sources, if
the current unification models are correct (Padovani \& Urry 1992),
giving a comoving space density for the host galaxies of $\sim
10^{-7}\Mpc^{-3}$, comparable to that of the brightest cD galaxies today.
The high energy density in the microwave background and the density of
the intergalactic medium may however seriously affect the size and
brightness of radio lobes (Dunlop \& Peacock 1992: this could explain the
lack of any resolved radio component to GB~1428+4217).

\section{Conclusions}

We report the results of the ASCA observations of the $z=4.72$ quasar
GB~1428+4217, the most distant X-ray source known. The observed energies
correspond to emission up to $\sim 57$ keV in the source rest frame.  The
flat X--ray spectrum, as well as the radio properties, indicate that the
source is a flat spectrum radio--loud blazar, dominated by relativistic
beaming from non--thermal plasma in a jet.  The slope of the X--ray
emission is among the flattest seen in blazars and suggests that
GB~1428+4217 is emitting most of its power in the MeV band.

The sparse spectral information does not allow us to constrain uniquely
the emission model. However it is important to note that plausible
models do not require parameters significantly different from
those of similar low redshift blazars.

Its properties are also similar to those of another radio--loud quasar at
$z\approxgt 4$, GB~1508+5714 (Moran \& Helfand 1997). In both sources the
X--ray emission dominates the (observed) radiative output and they show
no sign of absorption in excess to the Galactic one, contrasting with
that seen in the high $z$ radio--loud quasar sample presented by Cappi et
al. (1997).  Although two sources do not allow us to derive statistically
significant results, there is so far no sign of evolution either in the
spectrum or other properties for this class of objects. Low abundances in
such young objects could of course contribute to the lack of any
detectable absorption or emission features in their X-ray spectra.

Clearly, further information on the SED (in particular, the mm and MeV
spectral bands) and especially on the variability properties of
GB~1428+4217 will allow us to further constrain the models. It should be
noted that due to cosmological time dilation, variability timescales on
the shortest timescale (probably weeks) could be temporally resolved.

General statistical arguments predict that a significant source
population could be present at redshift $\approxgt 4$, with masses in
excess of $10^7 M_{\sun}$. 

It is remarkable that such a young object as GB~1428+4217 has very
similar properties to lower redshift sources. This emphasises that the
formation of relativistic jets is a ubiquitous phenomenon.

\section*{Acknowledgements}

We thank Gabriele Ghisellini for the use of the code for the homogeneous
emission model. The limits to the IRAS fluxes were obtained from SCANPI
at IPAC. The Royal Society (ACF, RGM), PPARC (KI, AC), the Smithsonian
Institution (WNB), and the Italian MURST (AC) are thanked for financial
support.

\end{document}